\documentclass[twocolumn,showpacs,preprintnumbers,amsmath,amssymb]{revtex4}
\usepackage{graphicx}
\usepackage{dcolumn}
\usepackage{bm}

\def\ket#1{\lvert#1\rangle}
\def\bra#1{\langle #1\lvert}

\def\ketbra#1#2{\lvert#1\rangle\langle#2\lvert}
\def\hc{{\ensuremath{\text{\,\it h.\,c.}}}}

\begin{document}


\title{Coherent energy manipulation in single-neutron interferometry}
\author{S. Sponar$^1$}
\author{J. Klepp$^{1,2}$}
\author{R. Loidl$^1$}
\author{S. Filipp$^1$}
\author{G. Badurek$^1$}
\author{Y. Hasegawa$^{1,3}$}
\author{H. Rauch$^1$}
\affiliation{%
$^1$Atominstitut der \"{O}sterreichischen Universit\"{a}ten, 1020 Vienna, Austria\\ $^2$Institut Laue-Langevin, B.P. 156, F-38042 Grenoble Cedex
9, France\\$^3$PRESTO, Japan Science and Technology Agency (JST), Kawaguchi, Saitama, Japan   }

\date{\today}
\begin{abstract}

We have observed the stationary interference oscillations of a triple-entangled neutron state in an interferometric experiment. Time-dependent
interaction with two radio-frequency (rf) fields enables coherent manipulation of an energy degree of freedom in a single neutron. The system is
characterized by a multiply entangled state governed by a Jaynes-Cummings Hamiltonian. The experimental results confirm coherence of the
manipulation as well as the validity of the description.
\end{abstract}


\maketitle

Since the pioneering work of Einstein, Podolsky, and Rosen \cite{EPR35} numerous experiments have exploited the concept of Nonlocality which
tests local hidden variable theories (LHVTs). The LHVTs are a subset of a larger class of hidden-variable theories namely the noncontextual
hidden-variable theories (NCHVTs). Noncontextuality implies that the value of a measurement is independent of the experimental context, i.e. of
previous or simultaneous measurements  \cite{Bell66,Mermin93}. Noncontextuality is a more stringent demand than locality because it requires
mutual independence of the results for commuting observables even if there is no spacelike separation \cite{Simon00}.

In the case of neutron experiments, entanglement is not achieved between particles, but between different degrees of freedom. Since the
observables in different Hilbert spaces commute with each other, the single neutron system is suitable for studying NCHVTs. Single-particle
entanglement, between the spinor and the spatial part of the neutron wave function \cite{Hasegawa03Bell}, as well as full tomographic state
analyses \cite{hasegawa2007tomography}, have already been accomplished. In addition, the contextual nature of quantum theory
\cite{Hasegawa2006contextual} has been demonstrated using neutron interferometry \cite{Rauch00Book}. Aiming at the preparation of a
single-particle multiply entangled state, implementation of another degree of freedom to be entangled with the neutron\char39{}s spin and path
degrees of freedom was a challenge.

The neutron\char39{}s energy seems to be an almost ideal candidate
for this third degree of freedom, due to its experimental
accessibility within a magnetic resonance field \cite{Badurek83TimeDepend}. For this purpose the
time evolution of the system is described by a photon-neutron
state vector, which is an eigenvector of the corresponding
modified Jaynes-Cummings (J-C) Hamiltonian
\cite{J-C63,Shore93JCM}. The J-C Hamiltonian can be adopted for a
system consisting of a neutron coupled to a quantized rf-field
\cite{Muskar87DressedNeutrons}.

This letter reports on observation of stationary interference patterns, confirming coherent energy manipulation of the neutron wavefunction.
This technique provides realization of triple-entanglement between the neutron\char39{}s path, spin and energy degrees of freedom.

Since  two rf-fields, operating at frequencies $\omega$ and $\omega/2$, are involved in the actual experiment, the modified corresponding J-C
Hamiltonian is denoted as
\begin{equation*}
\mathcal H_{\textrm{J-C}}=-\frac{\hbar^2}{2m}\nabla^2-\mu
B_0(\textbf{r}) \sigma_z+\hbar(\omega a_{\omega}^\dagger
a_{\omega}+\frac{\omega}{2} a_{\omega/2}^\dagger a_{\omega/2})
\end{equation*}
\begin{equation}\label{eq:J-CHamiltonian}
\!\!\!\!\!+\mu\Bigg(\frac{B^{(\omega)}_1(\textbf{r})}{\sqrt{N_{\omega}}}(
a_{\omega}^\dagger\widetilde\sigma+\hc)+\frac{B^{(\omega/2)}_1(\textbf{r})}{\sqrt{N_{\omega/2}}}(
a_{\omega/2}^\dagger\widetilde\sigma+\hc)\Bigg).
\end{equation}
with $\widetilde\sigma=\frac{1}{2}(\sigma_x+ i\sigma_y)$. The first term accounts for the kinetic energy of the neutron. The second term leads
to the usual Zeeman splitting of $2\lvert\mu\lvert B_0$. The third term adds the photon energy of the oscillating fields of frequencies $\omega$
and $\omega/2$, by use of the creation and annihilation operators $a^\dagger$ and $a$. Finally, the last term represents the coupling between
photons and the neutron, where $N_{\omega_j}=\langle a_{\omega_j} ^\dagger a_{\omega_j}\rangle$ represents the mean number of photons with frequencies $\omega_j$ in the rf-field.
Note that the first two and the last terms concern the spatial $\ket{\psi(\mathbf{r})}$ and the (time-dependent) energy $\ket{E(t)}$ subspaces of neutrons, respectively \cite{Subspaces}.

\begin{figure}[t]
{\includegraphics [width=90mm] {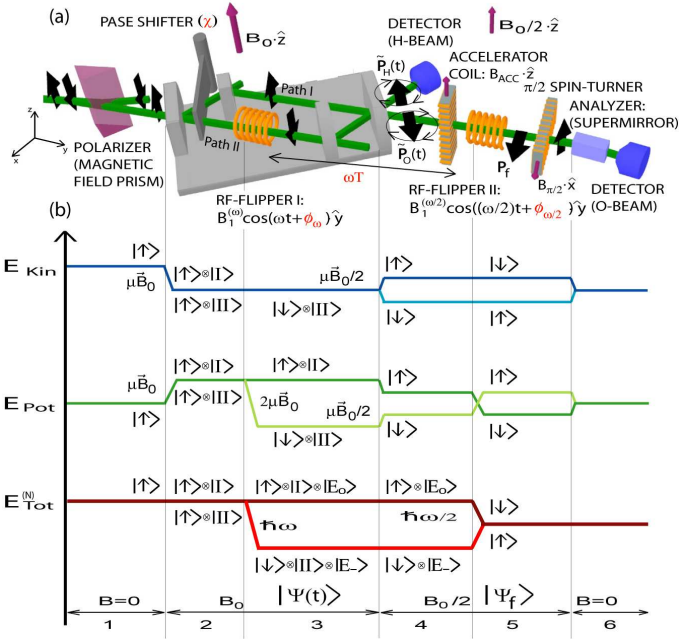}} \caption {({\it{a}}) Schematic view of the experimental setup for stationary observation of
interference between two rf-fields. Showing the arrangement of two radio-frequency flip coils (the first within one path of the skew-symmetric
Mach-Zehnder-type neutron interferometer and the other driven by the half frequency behind the interferometer), accelerator coil and $\pi/2$
spin-turner. Appropriate spin analysis of the neutron beam allows measurement of relative phase shifts. ({\it{b}}) Energy level diagram of the
two interfering sub-beams $\ket{\textrm{I}}$, $\ket{\textrm{II}}$ during their passage through the different static field regions
($\textrm{B}_0$,$\textrm{B}_0/2$, and B=0), including corresponding spin states $\ket{\uparrow}$, $\ket{\downarrow}$ and taking into account the
spin flips at rf-frequencies $\omega$ and $\omega/2$.}\label{fig:setup}
\end{figure}

The state vectors of the oscillating fields are represented by
coherent states $\ket{\alpha}$, which are eigenstates of
$a^\dagger$ and $a$. The eigenvalues of coherent states are
complex numbers, so one can write
$a\ket{\alpha}=\alpha\ket{\alpha}=\lvert\alpha\lvert
e^{i\phi}\ket{\alpha}\textrm{ with } \lvert\alpha\lvert=\sqrt{N}$.
Using Eq.\,(\ref{eq:J-CHamiltonian}) one can define a total state
vector including not only the neutron system
$\ket{\Psi_{\textrm{N}}}$, but also the two quantized oscillating
magnetic fields:
$\ket{\Psi_\textrm{i}}=\ket{\alpha_\omega}\otimes\ket{\alpha_{\omega/2}}\otimes\ket{\Psi_{\textrm{N}}}$.
In a  perfect Si-crystal neutron interferometer the wavefunction behind the first plate, acting as a beam splitter, is a linear superposition of
the sub-beams belonging to the right ($\ket{\textrm{I}}$) and the left path ($\ket{\textrm{II}}$), which are laterally separated by several
centimeters. The sub-beams are superposed at the third crystal plate and the wave function in the forward direction then reads as
$\ket{\Psi_{\textrm{N}}}\propto\ket{\Psi^{\textrm{(I)}}_{\textrm{N}}}+ \ket{\Psi^{\textrm{(II)}}_{\textrm{N}}}$, where
$\ket{\Psi^{\textrm{(I)}}_{\textrm{N}}}$ and $\ket{\Psi^{\textrm{(II)}}_{\textrm{N}}}$ only differ by an adjustable phase factor $e^{i\chi}$
($\chi=N_{\textrm{ps}}b_c\lambda D$, with the thickness of the phase shifter plate $D$, the neutron wavelength $\lambda$, the  coherent
scattering length $b_c$ and the particle density $N_{\textrm{ps}}$ in the phase shifter plate). By rotating the plate, $\chi$ can be varied
systematically. This yields the well known intensity oscillations of the two beams emerging behind the interferometer, usually denoted as O- and
H-beam \cite{Rauch00Book}. A sketch of the setup, split up into regions numbered from 1 on the left to 6 on the right side, is depicted in
Fig.\,\ref{fig:setup}.

In our experiment, only the beam in path II is exposed to the rf-field of frequency $\omega$, resulting in a spin flip process in region 3. The
spin flip configuration of the first rf-field ensures an entanglement of spin and spatial degree of freedom of the neutron state
\cite{Hasegawa03Bell}. Interacting with a time-dependent magnetic field, the total energy of the neutron is no longer conserved after the
spin-flip \cite{Golub94,Gaehler87ZeroField,Rekveldt2004,Alefeld81,Summhammer95MultiPhotonObservation}. Photons of energy $\hbar\omega$ are
exchanged with the rf-field. This particular behavior of the neutron is described by the dressed-particle formalism
\cite{Muskar87DressedNeutrons,Summhammer93MultiPhoton}. Consequently the two sub-beams $\ket{\textrm{I}}$ and $\ket{\textrm{II}}$ now differ in
total energy (see Fig.\,\ref{fig:setup}(b)). Therefore the neutron state can be considered to consist of the three subsystems, namely the total
energy, path and spin degree of freedom. In principle, a spin-independent energy manipulation of neutrons is also possible: for instance, the
up- and the down-spin wavepackets, separated by a so-called longitudinal Stern-Gerlach effect\cite{Arend04,Alefeldlongitudinal81}, undergo
successive fast-activated DC-RF and RF-DC flippers respectively, resulting in a positive energy-shift.

\begin{figure}[tbp]
\scalebox{0.28}{\includegraphics{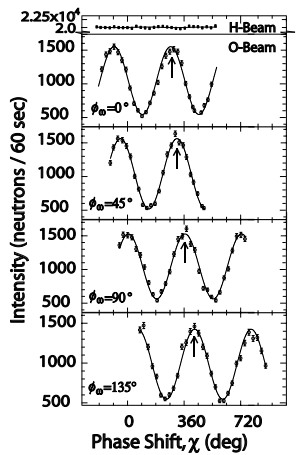}} \caption{\label{fig:graph1} Typical interference patterns of the H- and the O-beam. In the H-beam no
interference fringes are observed due to orthogonal spin states in the interfering sub-beams, whereas the O-beam exhibits time-independent
sinusoidal intensity oscillations, when the phase shifter plate ($\chi$) is rotated. A phase shift occurs on varying $\phi_{\omega}$.}
\end{figure}

A coherent superposition of $\ket{\textrm{I}}$ and $\ket{\textrm{II}}$ results in the multiply entangled dressed state vector, expressed as
\begin{equation*}
\ket{\Psi(t)}\propto\ket{\alpha_\omega}\otimes\ket{\alpha_{\omega/2}}\otimes\frac{1}{\sqrt{2}}\Big(\ket{\textrm{I}}\otimes\ket{E_0}\otimes\ket{\uparrow}
\end{equation*}
\begin{equation}\label{eq:EntangledState}
+e^{i\chi}\ket{\textrm{II}}\otimes e ^{i\omega t}\ket{E_0-\hbar\omega}\otimes e^{i\phi_\omega}\ket{\downarrow} \Big),
\end{equation}
where $\ket{\uparrow},\ket{\downarrow}$ denote the neutron\char39{}s up and down spin states referred to the chosen quantization axis. The state
vector of the neutron acquires a phase $\pm\phi_\omega$ during the interaction with the oscillating field, given by $B(t)=B_1 \cos(\omega t
+\phi_\omega)$, induced by the action of the operators $a_\omega$ and $a^\dagger_\omega$ in the last term of Eq.\,(\ref{eq:J-CHamiltonian}). The
neutron part of the total state vector is represented by a path-energy-spin entanglement within a single neutron system. At the last plate of
the interferometer (region 4) the two sub-beams are recombined, which is described by the projection operator $\hat
O^{\textrm{(P)}}=\frac{1}{2}\big(\ket{\textrm{I}}+\ket{\textrm{II}}\big)\big(\bra{\textrm{I}} +\bra{\textrm{II}}\big)$. Due to the orthogonality
of the energy and spin eigenstates the polarization is zero and no intensity modulations are observed in the H-beam, which is plotted in
Fig.\,\ref{fig:graph1}. A time-resolved measurement (see \cite{Badurek83TimeDepend}) can reveal the dynamic behavior of the polarization
expressed as
\begin{equation}\label{eq:Polarizationflippedrf}
\widetilde{\bf{P}}_{\textrm{O}}(t)=\Big(\cos\big(\chi-\omega t-\phi_\omega\big),\sin\big(\chi-\omega t-\phi_\omega\big),0\Big).
\end{equation}
This phenomenon has been measured separately  \cite{Badurek83TimeDepend}, and is related to the spinor precession known from zero-field spin-echo experiments \cite{Gaehler87ZeroField,Rekveldt2004}.

The beam recombination is followed by an interaction with the
second rf-field, with half frequency $\omega/2$, in region 5.
Mathematically the energy transfer is represented by the operator
$\hat
O^{\textrm{(E)}}=\frac{1}{\sqrt{2}}\ket{E_0-\hbar\omega/2}\big(\bra{E_0}+(\bra{E_0-\hbar\omega}\big)$,
respectively. The total state vector is given by
\begin{equation*}
\ket{\Psi_\textrm{f}}\propto\ket{\alpha_\omega}\otimes \ket{\alpha_{\omega/2}}\otimes \big(\ket{\textrm{I}}+\ket{\textrm{II}}\big) \otimes
\ket{E_0-\hbar\omega/2}
\end{equation*}
\begin{equation}\label{eq:FinalState}
\otimes\frac{1}{\sqrt{2}}\Big(e^{i\phi_{\omega/2}}\ket{\downarrow}+e^{i\omega T}e^{ i\chi}e^{i( \phi_\omega- \phi_{\omega/2})}\ket{\uparrow}\
\Big),
\end{equation}
where $\phi_\omega$ and $\phi_{\omega/2}$ are the phases induced
by the two rf-fields and $\omega T$ is the zero-field phase, with
$T$ being the neutron\char39{}s propagation time between the two
rf-flippers \cite{Sponar08}. The energy difference between the
orthogonal spin states is compensated by choosing a frequency of
$\omega/2$ for the second rf-flipper, resulting in a stationary
state vector. Hence the time dependence of the polarization vector
is eliminated:
\begin{equation}\label{eq:PolFinal}
{\bf P}_{\textrm{f}} =(\cos\Delta_{\textrm{tot}},\sin\Delta_{\textrm{tot}},0),
\end{equation}
where $\Delta_{\textrm{tot}}=(\chi-2\phi_{\omega/2}+\phi_{\omega}+\omega T)$, consists of the phases induced by the path (phase shifter $\chi$),
spin (phases of the two rf fields $\phi_{\omega},\phi_{\omega/2}$), and energy manipulation (zero-field phase $\omega T$). The principle of
energy compensation is visualized in Fig.\,\ref{fig:setup}(b). As seen from $\Delta_{\textrm{tot}}$ in Eq.(\ref{eq:PolFinal}) each of the three
degrees of freedom can be manipulated independently and the associated observables are separately measurable.

The arrangement of two rf-flippers of frequencies $\omega$ and $\omega/2$ can be interpreted as an interferometer-scheme for the
neutron\char39{}s total energy. Due to energy splitting the first rf-flipper generates a superposition of two coherent energy states, similar to
the action of the first beam-splitter of a Mach-Zehnder interferometer, where a single beam is split spatially into two coherent sub-beams. The
second flipper compensates the energy difference and therefore acts as a beam analyzer equivalent to the last beam-splitter of the
interferometer.

After applying a projection operator $\hat P^{(\textrm{S})}=\ketbra{\uparrow}{\uparrow}$ to the spin (region 6), the stationary interference oscillations
are given by $I_0\propto 1 +\nu \cos(\chi + \Phi+\omega T)$, introducing the fringe visibility $\nu$ and the relative phase $\Phi$. The relative
phase can be calculated as $\Phi=\phi_{\omega}- 2\phi_{\omega/2}$.
In the following experiment we demonstrate the coherence property of the modified J-C manipulation defined in Eq.\,(\ref{eq:J-CHamiltonian}) as
well as the phase dependence expressed above.

The experiment was carried out at the neutron interferometer instrument S18 at the high-flux reactor of the Institute Laue-Langevin in Grenoble,
France. A monochromatic beam, with mean wavelength $\lambda_0=1.91 \mbox{ \AA} (\Delta\lambda/\lambda_0\sim0.02$) and  5x5 mm$^2$ beam
cross-section, is polarized by a bi-refringent magnetic field prism in $\hat{\mathbf z}$-direction \cite{Badurek00FieldPrism}, see
Fig.\,\ref{fig:setup}(a) region 1. In a non-dispersive arrangement of the monochromator and the interferometer crystal the angular separation
can be used such that only the spin-up (or spin-down) component fulfils the Bragg-condition at the first interferometer plate (beam splitter) in
region 2. Behind the beam splitter the neutron\char39{}s wave function is found in a coherent superposition of
$\ket{\Psi^{\textrm{(I)}}_{\textrm{N}}}$ and $\ket{\Psi^{\textrm{(II)}}_{\textrm{N}}}$, and only $\ket{\Psi^{\textrm{(II)}}_{\textrm{N}}}$
passes the first rf-flipper mounted in one path of the interferometer. Acting like a typical NMR arrangement, rf-flippers require two magnetic
fields: A static field $B_0\cdot\hat{\mathbf z}$ with $B_0=\hbar \omega_{\textrm{rf}}/(2\lvert\mu\lvert)$ and a perpendicular oscillating field
$B_1^{(\omega)} \cos(\omega t +\phi_\omega)\cdot\hat{\mathbf y}$ with amplitude $B^{(\omega)}_1=\pi\hbar/(2\tau\lvert\mu\lvert$), where $\mu$ is
the magnetic moment of the neutron and $\tau$ is the time the neutron requires to traverse the rf-field region. The oscillating field is
produced by a water-cooled rf-coil with a length of 2\,cm, operating at a frequency of $\omega/2\pi=58$\,kHz. The static field is provided by
the uniform magnetic guide field $B_0\sim 2$\,mT, which is produced by a pair of water-cooled Helmholtz coils. However, outside the rf-coil the
Larmor precession around the static magnetic guide field induces an additional phase.

\begin{figure}[bp]
\scalebox{0.28}{\includegraphics{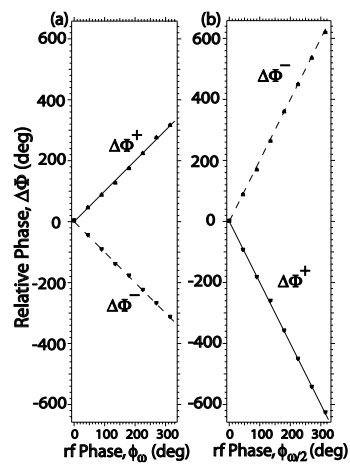}} \caption{\label{fig:graph3} Relative phase $\Delta\Phi^\pm$ vs. ({\it{a}}) $\phi_{\omega}$ and ({\it{b}}) $\phi_{\omega/2}$. The sign of the phase depends on the chosen initial polarization.}
\end{figure}
The two sub-beams are recombined at the third plate (region 4)
resulting in a time-dependent state vector due to the different
energies of the two partial wavefunctions. Since the two
superposed spin states are orthogonal, no intensity modulation is
observed, as seen at the H-detector. In contrast, the O-beam
(forward direction) passes the second rf-flipper, operating at
half the frequency of the first rf-flipper. The oscillating field
is denoted as $B^{(\omega/2)}_1 \cos\big((\omega/2) t
+\phi_{\omega/2}\big)\cdot\hat{\mathbf y}$, and the strength of
the guide field was tuned to about $1\,$mT in order to satisfy the
frequency resonance condition.

This flipper compensates the energy difference between the two spin components, by absorbtion and emission of photons of energy
$E=\hbar\omega/2$. The phases of the two guide fields and the zero-field phase  $\omega T$ were compensated by an additional Larmor precession
within a tunable accelerator coil with a static field, pointing in the $\hat{\mathbf z}$-direction. Finally, the spin is rotated back to the
$\hat{\mathbf z}$-direction by use of a $\pi/2$ static field spin-turner, and analyzed along the $\hat{\mathbf z}$-direction due to the spin
dependent reflection within a Co-Ti multi-layer supermirror. Typical interference patterns are depicted in Fig.\,\ref{fig:graph1}. In the O-beam
a fringe contrast of 52.4(2)\,\% is achieved, whereas no oscillation was observed in the H-detector, where no further manipulations were
applied.

It is possible to invert the initial polarization simply by rotating the interferometer by a few seconds of arc, thereby selecting the spin-down
component to enter the interferometer, which is expected to lead to an inversion of the relative phase. In order to observe a relative phase
shift, in practice it is necessary to perform a reference measurement. This is achieved by turning off the rf-flipper inside the interferometer,
thus yielding the relative phase difference $\Delta\Phi^\pm=\pm\phi_\omega\mp2\phi_{\omega/2}$, where $\pm$ denotes the respective initial spin
orientation. Figure \,\ref{fig:graph3}(a) shows a plot of the relative phase $\Delta\Phi^\pm$ versus $\phi_{\omega}$, with $\phi_{\omega/2}=0$,
and a phase shift $\Delta\Phi^\pm$ caused by a variation of $\phi_{\omega}$. As expected, the slope is positive for initial spin up
orientation(1.007(8)), and negative for the spin down case(-0.997(5)). In Fig.\,\ref{fig:graph3}(b) $\phi_{\omega/2}$ is varied, while
$\phi_{\omega}$ is kept constant, yielding slopes of -1.995(8) and  1.985(7), depending again on the initial beam polarization.

At this point the geometric nature of $\Delta\Phi^\pm$ should be emphasized. Within the rf-flipper that is placed inside the interferometer, the
neutron spin traces a semi-great circle from $\ket{\uparrow}$ to $\ket{\downarrow}$ on the Bloch sphere and returns to its initial state
$\ket{\uparrow}$ when passing the second rf-flipper. This procedure is repeated along different semi-great circles when varying $\phi_{\omega}$
or $\phi_{\omega/2}$ respectively. The two semi-great circles enclose an angle $\phi_\omega-\phi_{\omega/2}$ and hence a solid angle
$\Omega=2(\phi_\omega-\phi_{\omega/2})$. The solid angle $\Omega$ yields a pure geometric phase $\Phi^\pm_{\textrm{G}}=\Omega/2$ as in
\cite{Badurek2000GeoPhaseSep,GeoPhase}.


Our work can be seen within a framework related to tripartite entanglement. There are two non-equivalent classes of tripartite entanglement
represented by the Greenberger-Horne-Zeilinger (GHZ) state \cite{GHZ89Pro,GHZ90} and the W state \cite{Dur00} when the three quantum subsystems
have non-local correlations. Classification of a GHZ-like state in a single neutron system will be the subject of forthcoming work. In addition,
we claim that preparation of other types of triple entanglement can be realized using neutron interferometry and spin precession. For instance
creation of a W state can be achieved with rf-flippers within a double loop interferometer. It is worth noting, that the operation of the
rf-flipper within the interferometer could be interpreted as a "CNOTNOT-gate", with path as control qubit and energy and spin as target qubits.

In summary, we have established a technique of coherent energy manipulation, by utilizing the neutron interferometer in combination with two
rf-fields to observe time-independent interference patterns. Energy splitting provides an additional degree of freedom, available for multiple
entanglement of path, spin and energy of the neutron. Our data verify theoretical predictions and illustrate the significance of single particle
entanglement.

This work has been partly supported by the Austrian Science
Foundation, FWF (P17803-N02 and F1513). Y.H. would like to thank
the Japan Science and Technology Agency (JST) for financial
support.


\begin{thebibliography}{26}
\bibitem{Subspaces}
Factorization of wavefunctions is a common technique, e.g., to solve time-independent
Schr\"{o}dinger equations.
\end{thebibliography}
\end{document}